\documentstyle[12pt]{article}
\font\twlmsy=msbm10 at 12pt
\font\sevenmsy=msbm8
\font\fivemsy=msbm6
\newfam\Bbbfam
\textfont\Bbbfam=\twlmsy
\scriptfont\Bbbfam=\sevenmsy
\scriptscriptfont\Bbbfam=\fivemsy

\newcommand{\beq}{\begin{equation}}
\newcommand{\eeq}{\end{equation}}
\newcommand{\bea}{\begin{eqnarray}}
\newcommand{\eea}{\end{eqnarray}}
\newcommand{\beas}{\begin{eqnarray*}}
\newcommand{\eeas}{\end{eqnarray*}}
\newcommand{\beqs}{\begin{displaymath}}
\newcommand{\eeqs}{\end{displaymath}}

\newcommand{\ben}{\begin{equation}}
\newcommand{\een}{\end{equation}}

\newcommand{\bdm}{\begin{displaymath}}
\newcommand{\edm}{\end{displaymath}}

\begin{document}
\topmargin 0pt
\oddsidemargin 5mm
\headheight 0pt
\topskip 0mm


\pagestyle{empty}

\hfill RH-08-96

\vspace{2.5 truecm}

\begin{center}

{\Large \bf Scaling and Correlation Functions 

in a Model of a Two-dimensional 

Earthquake Fault}



\vspace{0.6 truecm}

{\em Thordur Jonsson\footnote{e-mail: thjons@raunvis.hi.is}\\ Sigurdur
Freyr Marinosson\footnote{e-mail: sigurdm@raunvis.hi.is} }

\bigskip

Raunvisindastofnun Haskolans, University of Iceland \\
Dunhaga 3, 107 Reykjavik \\
Iceland

\vspace{1.2 truecm}

\end{center}

\noindent
{\bf Abstract.} We study numerically a two-dimensional version of the
Burridge--Knopoff model.  We calculate spatial and temporal correlation
functions and compare their behaviour with the results found for the
one-dimensional model.  The Gutenberg--Richter law is only obtained for
special choices of parameters in the relaxation function.  
We find that the distribution of the fractal dimension of the 
slip zone exhibits two well-defined peaks corresponding to intermediate
size and
large events.

\bigskip


\vfill

\newpage
\pagestyle{plain}

\section{Introduction}
A mathematical model of an earthquake fault, originally introduced by
Burridge and Knopoff \cite{bk}, has been studied extensively in recent
years, see e.g.\ \cite{cl4,cl1,nakanishi2,olami1}, 
as an example of a self-organized critical system \cite{bak1,bak2}.
The motivation has also been to understand better the mechanism of
earthquake generation as well as the search for methods to predict
earthquakes \cite{cl3,pepke}.  
The quantities of main interest have been the scaling law
(Gutenberg--Richter law \cite{gutenberg}) relating the 
frequency of earthquakes to their size, other scaling laws 
as well as spatial and temporal correlations.

In \cite{cl2,cl3} the one-dimensional model is studied in detail by
integrating numerically the equations of motion.  In \cite{nakanishi} a
simplified model with qualitatively similar behaviour but simpler time
development, corresponding to coupled maps,  was introduced. 
This model, which is analogous to sandpile models \cite{sandpiles},  
was studied in more detail in \cite{nakanishi2,thordur}.  Some special
two-dimensional models are studied in \cite{olami1,olami3,olami2}, see 
\cite{old1,old2} for older work in the same vein. 
The purpose of this article is to generalize the study of \cite{thordur}
to two dimensions and compare the one and the two dimensional cases.

We begin by defining the model and the observables of interest. 
Then we describe
the results of extensive simulations of the size distribution, spatial
and temporal correlations as well as the structure of the slip zone.  
We briefly describe the algorithm,
which is essentially the one of \cite{grassberger}, and 
finally we compare our
results with previous work and discuss some outstanding questions.

\section{The model}
We describe here the two-dimensional version of the 
Burridge-Knopoff model and it will be clear from the discussion how to
generalize it to higher dimensions.  The higher dimensional analogue
does of course not have any straightforward physical interpretation.  
We restrict ourselves to the
homogeneous model.  The model 
can be described as a two-dimensional regular array 
of blocks of equal mass, which slide, subject to a velocity dependent
friction, on a plane which can be thought of as one side of an
earthquake fault.  The blocks are 
labelled by pairs of integers $(i,j)$, where $i=1,\ldots N_1$ and
$j=1,\ldots ,N_2$, such that $x_{i,j}$ is the coordinate of the block
labelled by $(i,j)$.
Each block is connected to its nearest neighbours in the first
coordinate direction by a spring with spring constant $k$ and to its
nearest neighbours in the second coordinate direction by leaf springs
with a spring constant $k'$.  
We only consider the isotropic case
where $k'=k$.  
In addition, each block is connected by a leaf spring, with spring
constant $k''$, to a sliding plane which lies parallel to the plane of the
blocks and moves with velocity $v_p$.   This plane can
be thought of as the other side of the fault.  
For a discussion of 
boundary conditions we refer to \cite{olami2}.  
 For convenience we choose units
such that $k''=1$ and $v_p=1$.

If we start the system at rest in a configuration where
the elastic force on each block is smaller than the static friction force, then
the blocks remain at rest until the elastic force on one of the blocks exceeds
a threshold value $f_{th}$ which equals the static friction.  Then 
this block starts to slide,  with increasing
velocity to begin with, 
since the friction is a decreasing function of velocity for
small velocities in physically realistic models, see \cite{friction}.  The
motion of the first block may cause the force on some of 
the neighbouring blocks to
exceed threshold so they begin sliding and so on.  In some cases the
motion of the first block unleashes an avalanche of sliding blocks and
this is regarded as a large earthquake.  Eventually all the blocks come
to rest again, at least if $v_p$ is small enough compared to the natural
velocities in the spring-block system.

We normalize the coordinates such that 
$x_{i,j}$ is the displacement of the $(i,j)$ block from its 
equilibrium position at time $t=0$.
The total nonfrictional force $f_{i,j}$ 
on this block at time $t$ due to the springs is then given by
\beq
f_{i,j}=-(x_{i,j}-t)+k\sum _{(n,m)}(x_{n,m}-x_{i,j}), \label{force}
\eeq
where the sum runs over the four nearest neighbours of the site $(i,j)$.
We do not study the Newtonian equations of motion for this system (see
\cite{cl1}) but rather use the simpler discrete 
dynamics (\cite{nakanishi,olami1})
which we now describe.   The elastic force increases linearly with
time on each block, so there is generically a unique 
first block which reaches the threshold for moving.  We let this block
move along the first coordinate axis (the direction of the driving
force) until the elastic force on it has reached the value
\beq
f'_{i,j}=\phi (f_{i,j}),
\eeq
where $\phi$ is a decreasing function which describes
the force relaxation in the system.  This motion is assumed to take
place in zero time
and meanwhile all the other blocks remain at rest.  
For earthquake dynamics this is an excellent approximation.
The slip of the $(i,j)$ block gives
rise to an increase in the force on its four nearest 
neighbours by an amount
$
{1\over 4}\Delta (f_{i,j}-\phi (f_{i,j})),
$
where  
\beq
\Delta = {4k\over 1+4k}.
\eeq
For a $d$-dimensional system $4$ is replaced by $2d$.
In general this force relaxation is modified at the boundary.
The parameter $\Delta$ is a measure of the stiffness of the system or,
if one prefers, the amount of conservation of elastic stress
in the system, $\Delta =1$ corresponding to a conservative system.

In practice we take $\phi$ so that
\beq
\phi (f_{th})=f_{th}-\delta\! f,
\eeq
where $\delta\! f>0$ is the smallest possible force relaxation.
By making one more choice of units we may assume that $f_{th}=1$.
Using the linear relationship between the coordinates of the blocks 
and the elastic 
forces we can easily compute the slips of blocks during earthquakes. 

If the force relaxation 
function $\phi (f)$ decreases fast for $f$ close to $f_{th}$, then 
a small slip is easily amplified as we discuss in more detail below.  
In our simulations we follow \cite{cl1,nakanishi} and use  
$\phi =\psi _{\alpha}$, where 
\beq
\psi _{\alpha}(x)={(2f_{th}-\delta\! f)^2\over
\alpha (x-f_{th})+(2f_{th}-\delta\! f)}-f_{th}
\eeq
and $\alpha >0$.  Note that $\psi _{\alpha}'(f_{th})=-\alpha$ so the size of 
$\alpha$ is a measure of how easily small slips can trigger large ones.
We also discuss briefly the case $\phi =0$ considered in \cite{olami1,olami2}
and in more detail in \cite{grassberger}.

\section{Observables}
We now describe the quantities that have been of main
interest in the study of the model.  In addition we introduce a natural 
notion
of a Hausdorff dimension for the slip zone.

If $x_{i,j}$ and $x'_{i,j}$ are the coordinates of the blocks before and
after an earthquake, its
moment is defined as
\beq
M=\sum _{i=1}^{N_1} \sum _{j=1}^{N_2} (x_{i,j}'-x_{i,j}),
\eeq
There are two other quantities of interest in describing the size of an
earthquake, the number of participating blocks which we shall denote by
$N$ and the energy $E$ realeased which is obtained by comparing the
potential energy in the springs before and after an event.

The magnitude of an earthquake
is defined as the logarithm of its ``size" so different notions of ``size"
give rise to different notions of magnitude.  In particular, we shall see
that the moment magnitude $\mu =\log M$, 
does not obey a Gutenberg--Richter law in two
dimensions except for very special choices of the parameters.

Before proceeding let us discuss qualitatively the main features of the
events observed in the case $\phi =\psi _\alpha$.
For a soft system the blocks move independently and only make the
minimal jumps corresponding to the force relaxation $\delta f$.   
In fact, the system is periodic when it is
sufficiently soft; if a given block moves at a certain time, then all
the other blocks move before the given block moves again.
At a sharply defined value of $\Delta =\Delta _0$ 
 events with more than one block begin to occur
in the system and there is no upper bound on the size of events
that can occur, see Fig.\ 1, where this scenario is illustrated for the
one-dimensional system.   The precise value of $\Delta _0$ depends
on $\alpha$ and the transition is sharper the larger the value of
$\alpha$.   In fact our data are 
consistent with a discontinous change in the size of 
the largest events for $\alpha >3$.  
We interpret this transition as being due to the
onset of
instability in the periodic state.  Once this state becomes unstable
and the motion of a block can trigger a motion of its neighbours, 
there is little that prevents earthquakes from spreading over the entire
system.  As we increase $\Delta$ further the dynamics becomes more
chaotic \cite{mogens}, the state 
of the blocks more irregular and earthquake fronts
are more likely to get trapped in  ``valleys" 
in the spring-block system.  As we increase $\Delta$ still further the
size of the maximal events passes through a minimum and eventually the
system becomes so stiff that almost any motion of a block can trigger a
systemwide slip.  In Fig.\ 2  we illustrate the same phenomenon 
for the two-dimensional system. 

Let us discuss this in more detail.  The displacement of the first block
to move in an earthquake is always
\beq
\theta_1=\delta\! f (1-\Delta ).
\eeq
Assuming that all blocks in an infinite 
system are at threshold the displacement 
of the $i$-th block in an earthquake is inductively given by 
\beq
\theta _i=1-\Delta +{\Delta\over 2}\theta _{i-1}
-(1-\Delta )\phi \left(1+{\Delta\theta _{i-1}\over 
2(1-\Delta )}\right)\label{theta}.
\eeq
This result was first derived for the one-dimensional system in 
\cite{thordur} but is in fact independent of the dimensionality of the system.
If $\phi$ is a decreasing function, then 
the sequence $\theta _{i}$ is easily seen to
be increasing and bounded so it converges 
to a value $\theta ^*$ which is the maximal slip of a single block
in an earthquake.  

Let us study the stability of the periodic state in a soft system 
more closely.
For simplicity we discuss the one-dimensional case but the mechanism is
the same in two and higher dimensions.
We assume that the system is in a state
where all events are minimal. 
Then we have for all $i$,
\beq
1- {\it \delta f} \leq f_i \leq 1,
\eeq
where $f_i$ is the total force from the springs on block $i$.
Let us assume that block $1$ is going to move next and $f_1=1$.
We now also put $f_2$ to threshold and we want to derive a sufficient
condition on $\alpha$ in $\psi_\alpha$ and $\Delta$ so that block $3$ 
will also move.
We have 
\beq
f_1 \mapsto 1-{\it \delta f}
\eeq
and 
\beq
f_2  \mapsto f'_2 = 1+\frac{1}{2}\Delta{\it \delta f}.
\eeq
Since
\beq
1-{\it \delta f} \leq f_3 \leq 1
\eeq
a sufficient condition for the
movement of the $3$rd block is
\beq
\frac{1}{2}\Delta(1-\phi(1+\frac{1}{2}\Delta {\it \delta f})) \geq
{\it \delta f}.
\eeq
With $\phi =\psi _\alpha$ this condition is equivalent to
\beq
\frac{(2-{\it \delta f})^2}{{\it \delta f}
(1-\frac{{\it \delta f}}{\Delta})}-\frac{4}{{\it \delta f}}+2 \leq
\alpha \Delta .
\eeq
Since $\delta f\ll 1$ it follows that the above inequality is roughly
equivalent to 
\beq
4\left(\frac{1}{\Delta}-1\right)+2 \leq \alpha \Delta
\eeq
which agrees quite well with the location of the onset of the
instability in the $\alpha \Delta$ plane, cf. Fig.\ 1.

We now introduce the correlation functions.
Let $F$ be a function of time and 
the coordinates $x_{i,j}$ of the
blocks.  We refer to such functions as observables.
The average $<F>$ of an observable $F$ is defined as
\beq
<F>=\lim _{T\to\infty}{1\over T}\int _0^TF \,dt,
\eeq
provided the limit exists.
In the simulations we choose of course a finite value of
$T$ which must be taken sufficiently large,
depending on the observable under study
as well as the system size.

We define 
\beq
\xi _{i,j}(t)=x_{i,j}(t)-v_pt.
\eeq
The average $<\xi _{i,j}>$ is well defined since the long time 
average of $x_{i,j}$ is proportional to $t$ with coefficient $v_p$.

The spatial correlation  function $G(i)$  is defined as
\beq
G(i)=<\xi _{i+N_1/2,N_2/2}\xi _{N_1/2,N_2/2}>-<\xi _{i+N_1/2,N_2/2}>
<\xi _{N_1/2,N_2/2}>.
\eeq
Here we have assumed that $N_1$ and $N_2$ are even numbers and chosen the 
center of the system as a reference point in order to
minimize finite size effects in the case of free or open boundary
conditions.  In the case of periodic boundary conditions it does of course
not matter which subtraction point we choose.   A more general
spatial correlation function is obtained by separating the two $\xi$
variables by an arbitrary lattice vector.  This more general correlation
respects the symmetry of the lattice but does not reveal any
features different from those of $G(i)$.

We define the time correlation function (or autocorrelation function) 
$C(t)$ as
\beq
C(t)=\lim _{T\to\infty}
{1\over T}\int _0^T\xi (t'+t)\xi (t')\, dt'-<\xi >^2,
\eeq  
where $\xi =\xi _{N_1/2,N_2/2}$.
Clearly $C(0)=G(0)$.

It is of interest to consider the shape of the slip zone in earthquakes.
 For this purpose we introduce the notion of the Hausdorff dimension
$\delta$ of the slip zone.  It is defined as
\beq
\delta ={\ln N\over \ln D}
\eeq
where $N$ is the number of blocks that participate in the event and $D$
is the length of the diagonal 
of the smallest rectangular box, with sides along the coordinate axes,  
which contains all the
blocks that participated in the earthquake.  This definition is
by no means unique but we find it convenient to work with.  All
reasonable definitions should be equivalent in the limit of large $N$
and $D$.

\section{Numerical results}
In this section we describe the results of simulations for the
relaxation function $\psi _3$.
Most of the results are for a $200\times 200$ block system with periodic
boundary conditions.
We begin by considering the size distribution of earthquakes.  We let
$R(\mu )$ denote the density of events per unit time with moment magnitude
$\mu$.  
There are three regimes in this distribution for all values of $\Delta$
and $\alpha$.  We discuss only the case of $\alpha =3$.  Other values of
$\alpha$ in the range from 2 to 5 do not lead to a qualitatively
diferent picture.

For small events the discretization shows clearly up.  In the intermediate
range from $\mu \approx -2$ to $\mu \approx 
0$ the distribution falls off but not linearly
except for $\Delta \approx 0.82$.  For smaller $\Delta$ the distribution
is concave in this region while it is convex for larger values of
$\Delta$, see Figs.\ 3-5.  
This is quite different from the one-dimensional model where the size
distribution is linear to a good approximation in the intermediate range
\cite{thordur}.
  The distribution of earthquakes as a function of the
energy is qualitatively similar.
However, the distribution as a function of the number of participating
blocks is linear to a very good approximation for
$\Delta \geq 0.8$ with a slope which varies continuously with $\Delta$,
see Figs.\ 6-7.  For smaller $\Delta$ the distribution becomes concave.

For $\Delta <0.65$ almost all events are minimal but around
$\Delta =0.65$ the instability discussed in the previous section sets in,
cf.\ Fig.\ 2,
and for $\Delta =0.67$ 
some events extend over the entire system with almost no events of
an intermediate size.  As $\Delta$ increases further 
events of intermediate size
begin to appear and finally for the very stiff system the peak at large
events begins to rise again.

The behaviour of the size distribution is also reflected in the
correlation functions.  Let us first discuss the spatial correlation
$G(i)$, see Fig.\ 8.  For small $\Delta$ we find 
a random function of small amplitude since the system is periodic.  
As events involving more than one block begin 
to occur we find a sharply peaked function 
at $0$ which
explodes at the ``phase transition" point $\Delta =0.67$ to a broad
function whose decay is not very well determined due to finite size effects
and the very large
fluctuations in the simulation at this value of $\Delta$.  For larger
values of $\Delta$ the functions narrows until the prevalence of large
earthquakes reaches a minimum at $\Delta\approx 0.82$.  It broadens
again as  $\Delta$ is increased further.   In an infinite system 
the correlation should decay monotonically to zero.  We see marked 
finite size effects for $\Delta =0.67$ as well as for $\Delta =0.95$.
This is in a qualitative
agreement with the behaviour of the spatial correlation in the
one-dimensional model and the value of $G(0)$ is roughly proportional to
the maximum displacement of blocks in a single earthquake, cf.\
\cite{thordur}. 

The autocorrelation function $C(t)$ is considerably more difficult to
calculate in two dimensions than in the one-dimensional case.  As far as
we can see it has qualitatively the same behaviour in one and two dimensions 
as long as the 
system is not too stiff.  The correlation decays rapidly with
well defined oscillations whose frequency is well explained by the
maximum slip of a single block and the driving velocity, see Fig.\ 9.
The data does not allow us to say anything about the decay of the envelope.
For a stiff system ($\Delta\approx 0.95$) the oscillations disappear,
a phenomenon not seen in the study of the one-dimensional system.

In the one-dimensional model the structure of the slip zone in an
earthquake is always trivial since it is connected.  
In two dimensions the situation is
quite different.   The structure of the slip zone is intimately
related to the structure of the displacement field $\xi _{i,j}$.
In Figs.\ 10 - 12 we indicate by cross those blocks which are subject to an
eleastic force greater than $0.9$ at a particular time.
The value $0.9$ is chosen solely for illustrative purposes.
We see that the
softest system is quite regular and the boundary between the set of blocks
which experience a force in excess of $0.9$ and the rest of the blocks
consistes mostly of straight lines.  As we increase $\Delta$ the system
becomes more irregular with valleys of all sizes.   The system becomes
more homogeneous as $\Delta$ is increased still further.   The shape of the
slip zone for typical
earthquakes  occuring in systems with these three values of the stiffness
parameter are illustrated in Figs.\ 13 - 15.  In the soft system the slip
zone has usually a regular boundary, mirroring the structure of the
underlying displacement field.  The slip zone on Fig. 13 is atypical in the
sense that most large earthquakes for $\Delta =0.67$ extend 
over a large fraction of the
system.  For intermediate values of $\Delta$ the slip zone becomes more
irregular.  For the very stiff system, large events tend to extend over most
of the system and the slip zone has an 
irregular boundary.   The Hausdorff dimension
corresponding to the slip zones in these three cases is clearly smallest for
the most chaotic system corresponding to 
the interemediate value $\Delta =0.82$ for the stiffness parameter.  

A further qualitative
insight into the dynamics of the system is obtained by following the
development of earthquakes in real time and see it move block by block.  In
the soft system the earthquake front propagates in the beginning along
ribbons which make a $45^\circ$ angle with the coordinate axes.  This is
expected because in this way two blocks can cooperate to topple a block
which they have as a common nearest neighbour.  As the slip zone grows in
size it often becomes roughly hexagonal for the largest events.
In the stiffest system we also observe the hexagonal expansion of the slip
zone but the ribbons are absent and earthquakes engulf the more stuck
regions.  The intermediate system is most irregular as expected. 

The average Hausdorff dimension is not the relevant measure
of the structure of the slip zone.  This is illustrated in Figs.\
16 - 17 for the system with $\Delta =0.67$.  Fig.\ 16 is a histogram of the
Hausdorff dimension of events with more than 100 blocks.  There are two
peaks in the distribution at the values $\delta =1.86$ and $\delta =1.52$.
If we look at Fig.\ 17, which shows
the distribution of events with more than 1000 blocks
participating,  we see that the peak at the
lower value of $\delta$ has disappeared.  The earthquakes in the $\Delta
=0.67$ system therefore 
fall into two classes: Those that extend over almost
the entire system and have a slip zone with straight edges and those that are
bounded in size.  These two classes of events correspond to the tail and
main body of
the $R-\mu$ distribution in Fig.\ 3.  The slip zone illustrated in Fig.\ 13
corresponds to the events that have a Hausdorff dimension close to 2, but
most events in this class are in fact much larger.

We observe a similar distribution of the Hausdorff dimension in the stiff
system $\Delta =0.95$ with large Hausdorff dimension for the large events
whereas the events in the intermediate system ($\Delta =0.82$) have a
Hausdorff dimension that peaks around $\delta =1.6$ and does not depend
sensitively on the size cutoff, see Figs.\ 18 - 21.

\section{The OFC model}
The model studied by by Olami, Feder and Christensen \cite{olami1}, see
also \cite{olami3,olami2,grassberger}, is the one we have discussed
above with the force relaxation function $\phi =0$.  For describing real
earthquake faults this is probably a highly unrealistic choice of force
relaxation since the typical stress drop in earthquakes is believed to
be of the order of 10\% or less \cite{scholz}.  
In this model the size distribution (defined as the logarithm of the number
of participating blocks) falls nicely on a straight line.  Other notions of
size do not yield a linear distribution.

In order to compare the OFC model with the one we have simulated we have
calculated the spatial as well as the temporal corelation functions, see 
Figs.\ 22 and 23.  They
are qualitatively quite similar to the ones found in the Nakanishi model
with the main difference that the autocorrelation function does not
oscillate around zero but around a positive value, see Fig.\ 23.  
This might very well 
be a finite size effect which are known to make the analysis of the
OFC model quite subtle \cite{grassberger}.

\section{The algorithm}

In order to simulate the temporal behaviour of an observable ${X}$
we have to run the system for a certain time $T$ in order to obtain
resonable statistics.  In most cases we are not interested in finite
size effects and want to have the system as large
as possible.  It is therefore important how the computational
time scales with $N$ when $T$ is constant.
For many parameter values the average size of an event is not a function
of $N$ and therefore the generation of an earthquake is an
$O(1)$ algorithm
and the average time between earthquakes scales like $N^{-1}$.
To run the dynamics in an $N$ block system for a time $T$ we therefore
need a number of earthquakes proportional to $TN$.
If we had an $O(1)$ method of
finding the block closest to threshold we could carry out the simulation 
in a time $O(N)$.  Using a method introduced by Grassberger 
\cite{grassberger} this possible
under the assumption that the distribution of the forces on the blocks
has no delta-function peaks.  Our method is quite similar.
We use $M$ boxes $B[0],B[1],..,B[M-1]$, $M \propto N$,
which can
store blocks in doubly-linked lists.  The advantage of having a doubly-linked
list is that the removal of a block from such a list is an $O(1)$
operation.  The indexing of the $B$'s is understood
modulo $M$.  We define $\epsilon = \frac{2}{M-1}$ and
$q_i = f_i - t$ for $i=1,2,..,N$,
where $t$ is the physical time.
In our implementation we also have two other boxes $A_1$ and $A_2$
which can hold blocks.  They are used to keep track of the blocks that
go over threshold.  We need two such boxes because we must keep track of
the order in which
the blocks move in an earthquake.
The idea is to keep the blocks in the $B$ boxes ordered in such a way
that the block closest to threshold can be found in a  time $O(1)$.

In the simulations reported in this paper we use a global relaxation
function $\phi$ and a global $\Delta$, corresponding to a homogeneous
system. With trivial modifications it is possible to use the
same algorithm where each block $i$ has its own relaxation function
$\phi_i$ satisfying $|\phi_i|<1$ and the springs between the blocks
have different spring constants.

In the beginning we set $t=0$ and
distribute the
$f_i$'s randomly $-1 < f_i < 1$.  We put block $i$ in
$B[\lfloor \frac{1- q_i}{\epsilon} \rfloor]$, where $\lfloor x\rfloor$
denotes the largest integer smaller than $x$.
The boxes $A_1$ and $A_2$ are empty. 
Then we perform the following steps consecutively:

\begin{enumerate}

\item 
All the blocks are experiencing force between $-1$ and $1$, $t \geq 0$, 
and for 
$i=1,2,..,N$ block $i$ is in $B[j_i]$ where $j_i =
\lfloor \frac{1- q_i}{\epsilon} \rfloor$.
This implies
\beq
-1-t < q_i \leq 1-t
\eeq
and then
\beq
\lfloor \frac{t}{\epsilon} \rfloor \leq j_i \leq  
\lfloor \frac{t}{\epsilon} \rfloor +M-1
\eeq
and
\beq
1+t-(j_i+1)\epsilon < f_i \leq 1+t-j_i\epsilon
\eeq
for all $i$.  From this it follows that if the box 
$B[\lfloor \frac{t}{\epsilon} \rfloor]$ is not empty, then it contains
the block $i_{\it max}$ closest to threshold.  If it is empty,
we look into the boxes $B[\lfloor \frac{t}{\epsilon} \rfloor +1]$,
$B[\lfloor \frac{t}{\epsilon} \rfloor +2]$,..
until we find a nonempty box, which then contains $i_{\it max}$.
We find the block closest to threshold and remove it from the box.
We define $f_{\it max} = f_{i_{\it max}}$ and 
${\it \delta t} = 1-f_{\it max}$.  We then 
increase $t$ by ${\it \delta t}$.
This does not affect the ordering in the boxes for the $q$'s
will not change.   After this $f_{\it max}$ is equal to $1$ and we
label the $i_{\it max}$ block $i_{\it active}$.

\item
We update the $i_{\it active}$ block according to the dynamics, i.e.
\beq
f_{i_{\it active}} \mapsto \phi(f_{i_{\it active}})
\eeq
and update $q_{i_{\it active}}$ accordingly.
Then we put it into the box 
$B[\lfloor \frac{1-q_{i_{\it active}} }{\epsilon} \rfloor]$ and 
remove all its neighbours $i_x$ from their boxes.  
Because we can adress them through their positions in the block-array 
this is an $O(1)$ operation.
We modify their $f$'s according to
\beq
f_{i_x} \mapsto f_{i_x}+\frac{1}{2D} \Delta (f_{i_{\it active}}-
\phi(f_{i_{\it active}})),
\eeq
where $D$ is the dimension of the system
and update their $q$'s.  If a neighbour becomes critical, i.e. 
has force above 1, we put it into the box $A_2$.  The rest is put back
in the $B$ boxes, $i_x$ in $B[\lfloor \frac{1-q_{i_x} }{\epsilon}
\rfloor]$.

\item
We look into the box $A_1$.  If it
is not empty we remove some block
$i$ from $A_1$, set $i_{\it active} = i$ and go back to step 2.
If the box $A_1$ is empty we go to the box $A_2$.  If $A_2$ is not 
empty we remove some block $i$ from it and set $i_{\it active} = i$,
move the rest of the blocks in $A_2$ to $A_1$, and go back to step 2.
If both $A_1$ and $A_2$ are empty, then there are no more critical blocks
at this time, i.e. the earthquake is over, and we go back to step 1.
   
\end{enumerate}
This completes our description of the algorithm.

\section{Discussion}
In this paper we have generalized the simulation of the Nakanishi model
to two dimensions.  We have not been able to go much 
beyond systems of size
$200\times 200$ so our results have considerable finite size effects which
are not straightforward to entangle.  In general these systems do not have
any translationally invariant thermodynamic limit as we saw clearly in the
case of a stiff one-dimensional system in \cite{thordur}.
The simulations reported here also show that when the system makes the
transition from the trivial periodic state when the blocks move
independently of each other there is a very abrupt change with systemwide
events, both in the one- and the two-dimensional models.

The only qualitative difference we have found between the one- and the
two-dimensional cases is the absence of any linear region in the $R-\mu$
distribution in two dimensions and the positivity of the autocorrelation
function for a stiff two-dimensional system.

We tried to find a transition from a one-dimensional behaviour, as one 
expects in a long and narrow system, to properly two dimensional behaviour.  
However, finite size effects did not allow us to see such a crossover
clearly.  In real earthquakes, the frequence size distribution is expected to
be governed by a two-dimensional theory for small events but by a 
one-dimensional theory for large events for which the seismogenic zone extends 
through the litosphere \cite{frequency}.  By a careful analysis of an 
elongated two-dimensional model we should be able to see such a transition 
in the model discussed here.

\bigskip

\noindent
{\bf Acknowledgement.} This work was supported by the Research Fund of the
University of Iceland, the Icelandic Science Foundation and Nyskopunarsjodur 
Namsmanna.

\newpage
\noindent
{\Large\bf Figure Caption}

\bigskip
\noindent
{\bf Fig. 1.} The figure illustrates, as a function of $\Delta$,
the average number of blocks participating
in the 10 largest out of $2\times 10^7$ events in a one dimensional system
consisting of 10,000 blocks for values of $\alpha$ ranging from 2 to 5.

\medskip
\noindent
{\bf Fig. 2.} The figure illustrates, as a function of $\Delta$, 
the average number of blocks participating  
in the 10 largest out of $10^8$ events in a $200\times 200$ block system
for $\alpha =3$.

\medskip
\noindent
{\bf Fig. 3.} The figure shows the relative frequency of earthquakes as
a function of their magnitude for $\Delta =0.67$.

\medskip
\noindent
{\bf Fig. 4.} The figure shows the relative frequency of earthquakes as
a function of their magnitude for $\Delta =0.82$.

\medskip
\noindent
{\bf Fig. 5.} The figure shows the relative frequency of earthquakes as
a function of their magnitude for $\Delta =0.95$.

\medskip
\noindent
{\bf Fig. 6.} The figure shows the relative frequency of earthquakes as
a function of the number of participating blocks for $\Delta =0.82$.

\medskip
\noindent
{\bf Fig. 7.} 
The figure shows the relative frequency of earthquakes as
a function of the number of participating blocks for $\Delta =0.95$.

\medskip
\noindent
{\bf Fig. 8.}
The spatial correlation function for 3 different values of $\Delta$.

\medskip
\noindent
{\bf Fig. 9.} The autocorrelation function for 3 different values of $\Delta$.

\medskip
\noindent
{\bf Fig. 10.} The crosses indicate those blocks that are subject to an
elastic force greater than $0.9$ in a $200\times 200$ block 
system with $\Delta
=0.67$.

\medskip
\noindent
{\bf Fig. 11.} The crosses indicate those blocks that are subject to an
elastic force greater than $0.9$ in a $200\times 200$ block 
system with $\Delta
=0.82$.

\medskip
\noindent
{\bf Fig. 12.} The crosses indicate those blocks that are subject to an
elastic force greater than $0.9$ in a $200\times 200$ block 
system with $\Delta
=0.95$.

\medskip
\noindent
{\bf Fig. 13.} The crosses indicate the position of the blocks that
participated in a typical earthquake in a $200\times 200$ block system with
$\Delta =0.67$.

\medskip
\noindent
{\bf Fig. 14.} The crosses indicate the position of the blocks that
participated in a typical earthquake in a $200\times 200$ block system with
$\Delta =0.82$.

\medskip
\noindent
{\bf Fig. 15.} The crosses indicate the position of the blocks that
participated in a typical earthquake in a $200\times 200$ block system with
$\Delta =0.95$.

\medskip
\noindent
{\bf Fig. 16.}  A histogram illustrating the distribution of the 
Hausdorff dimension of the slip zone for events consisting of more than 
100 blocks in a system with $\Delta =0.67$. 

\medskip
\noindent
{\bf Fig. 17.}  A histogram illustrating the distribution of the 
Hausdorff dimension of the slip zone for events consisting of more than 
1000 blocks in a system with $\Delta =0.67$.

\medskip
\noindent
{\bf Fig. 18.}  A histogram illustrating the distribution of the 
Hausdorff dimension of the slip zone for events consisting of more than 
100 blocks in a system with $\Delta =0.82$.

\medskip
\noindent
{\bf Fig. 19.}  A histogram illustrating the distribution of the 
Hausdorff dimension of the slip zone for events consisting of more than 
1000 blocks in a system with $\Delta =0.82$.

\medskip
\noindent
{\bf Fig. 20.}  A histogram illustrating the distribution of the 
Hausdorff dimension of the slip zone for events consisting of more than 
100 blocks in a system with $\Delta =0.95$.

\medskip
\noindent
{\bf Fig. 21.}  A histogram illustrating the distribution of the 
Hausdorff dimension of the slip zone for events consisting of more than 
1000 blocks in a system with $\Delta =0.95$.

\medskip
\noindent
{\bf Fig. 22.} The spatial correlation function in the OFC model with 
$\Delta = 0.8$.

\medskip
\noindent
{\bf Fig. 23.}
The autocorrelation function in the OFC model with $\Delta =0.8$.

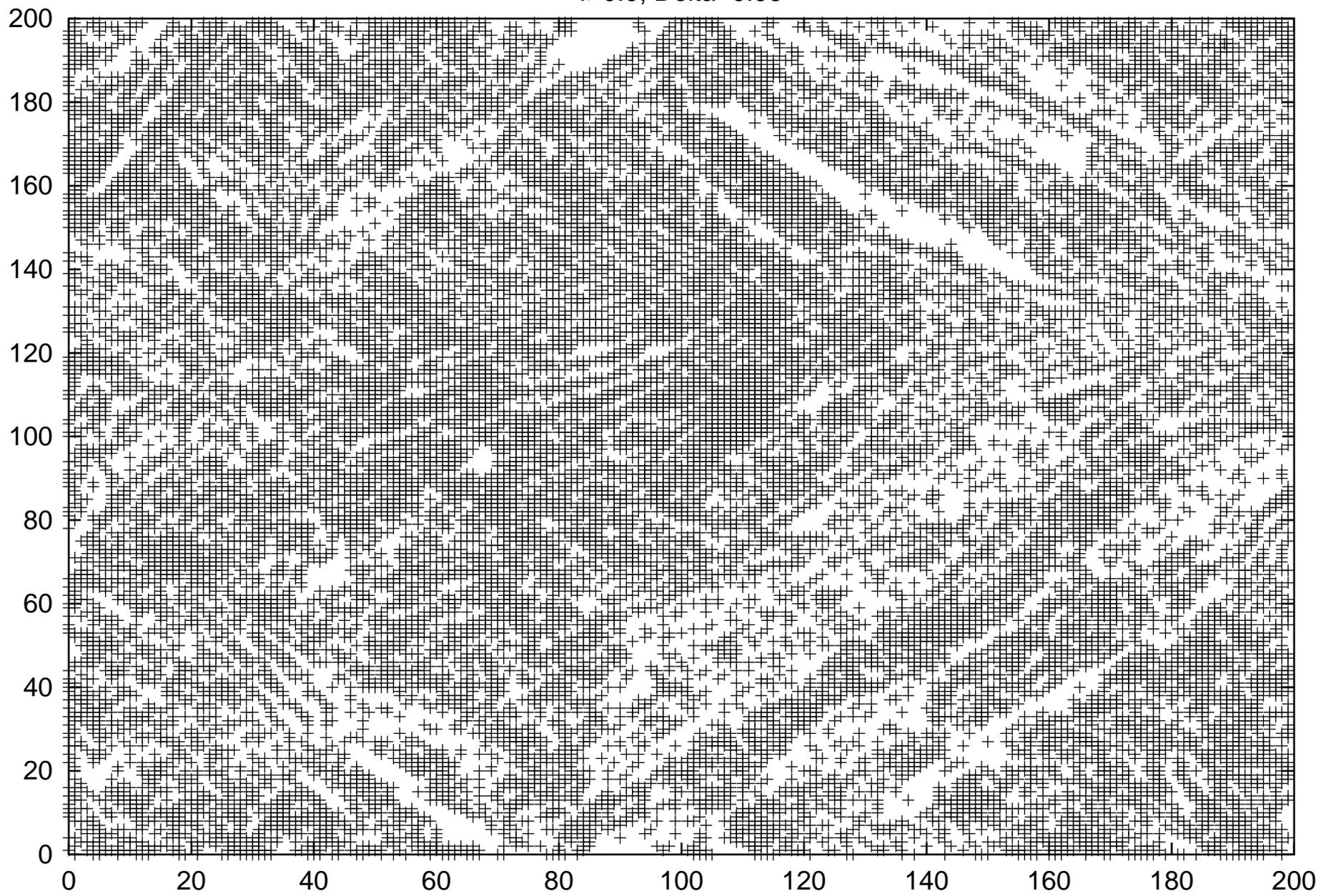
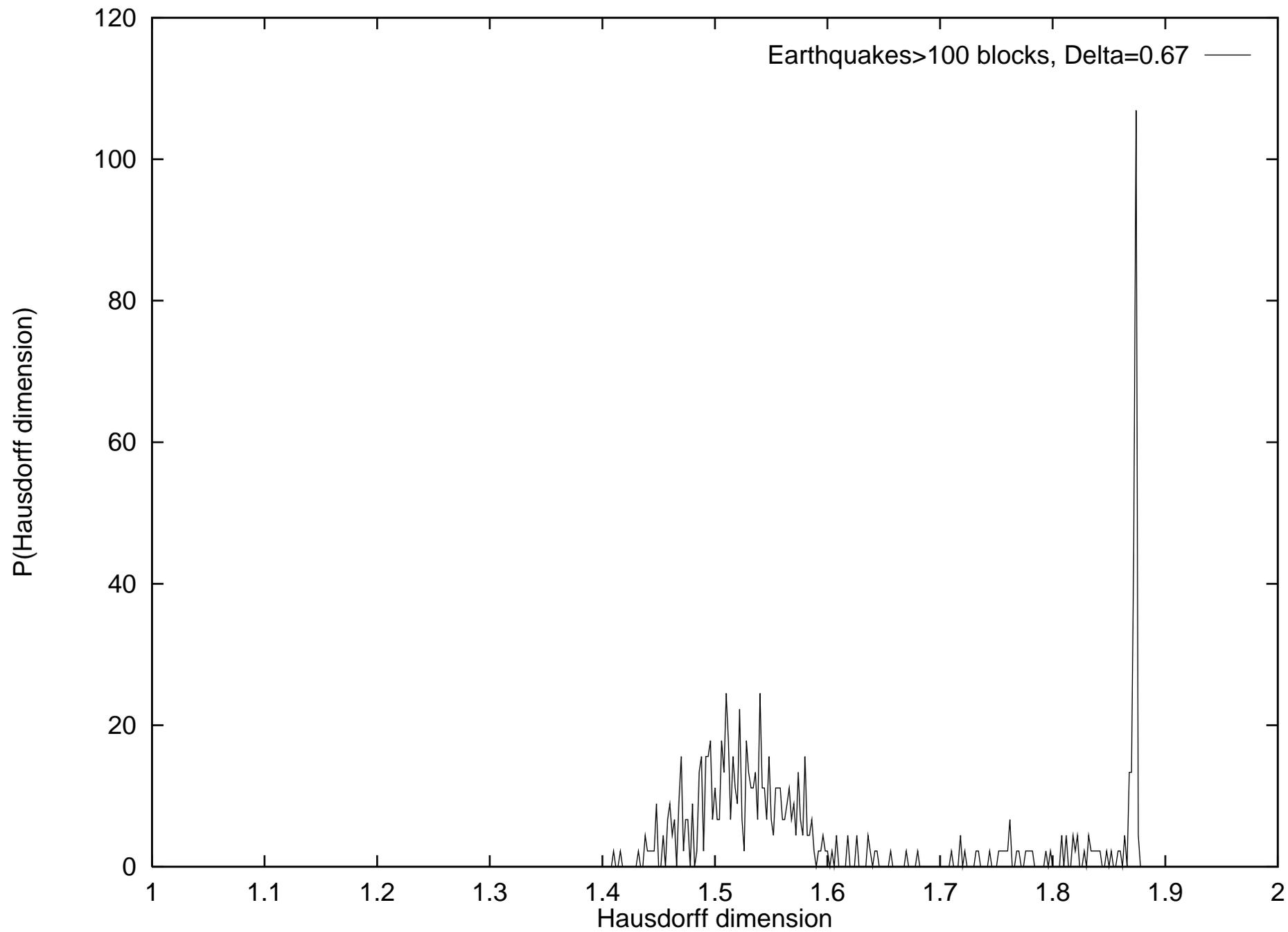
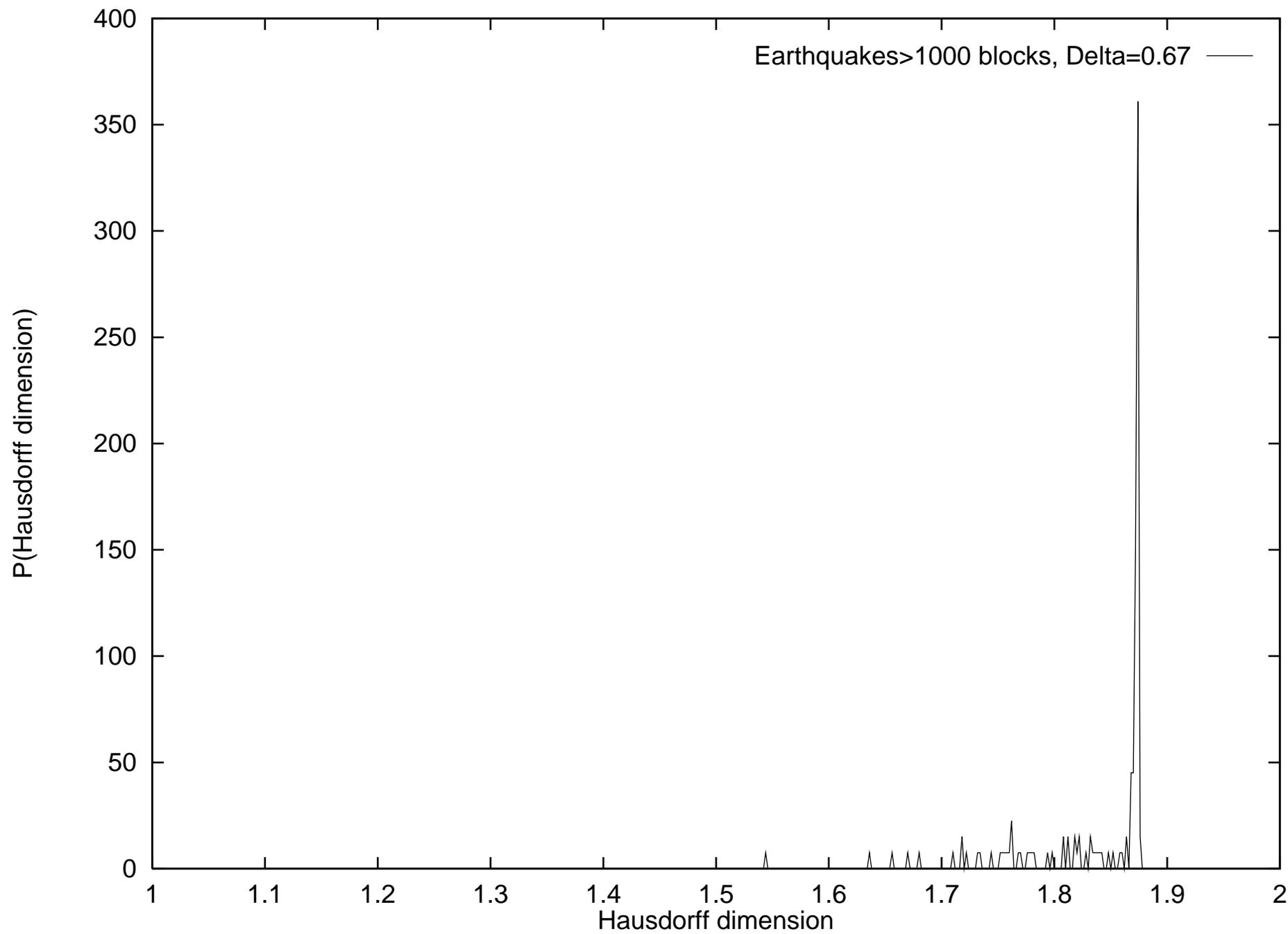
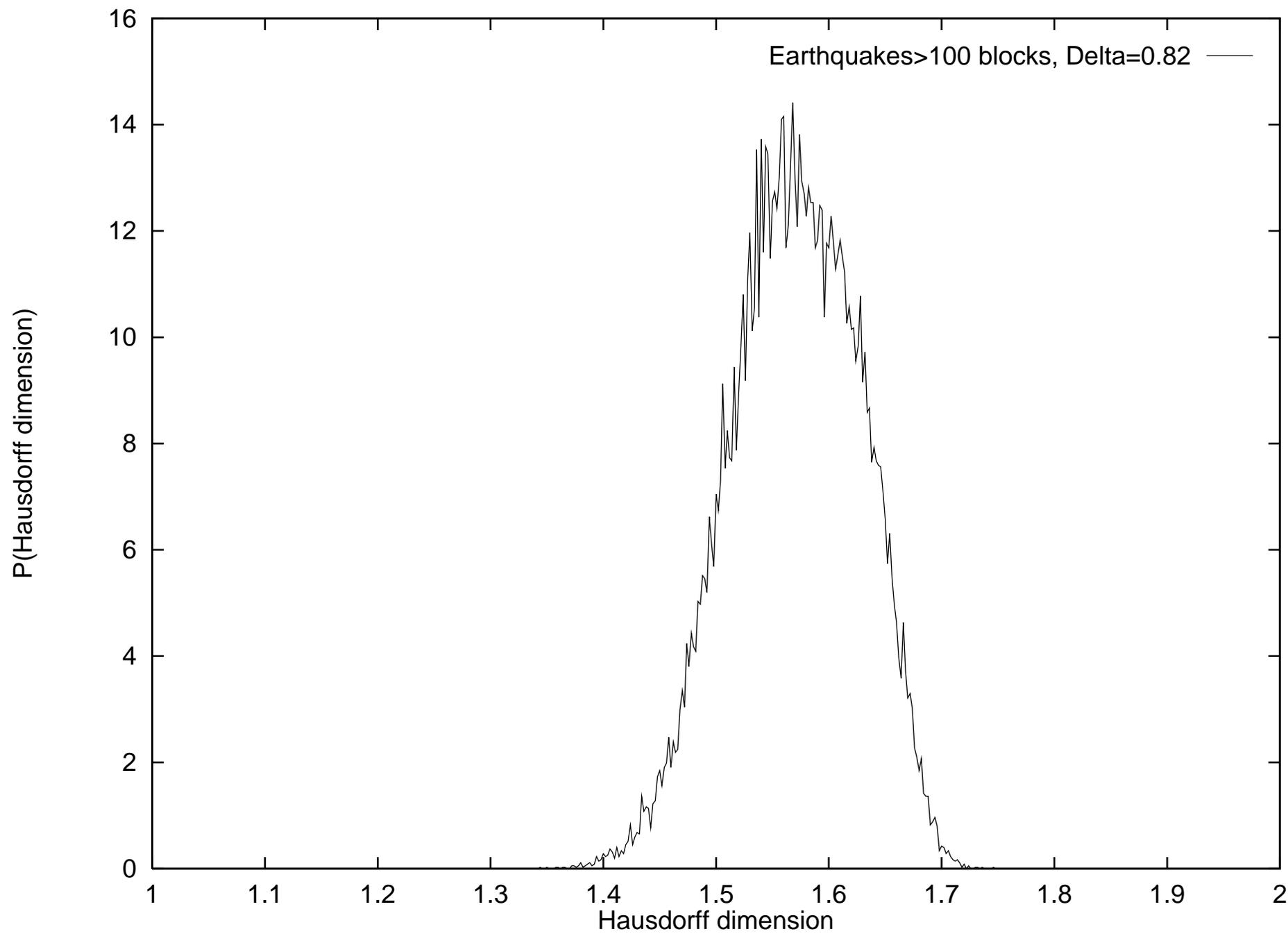
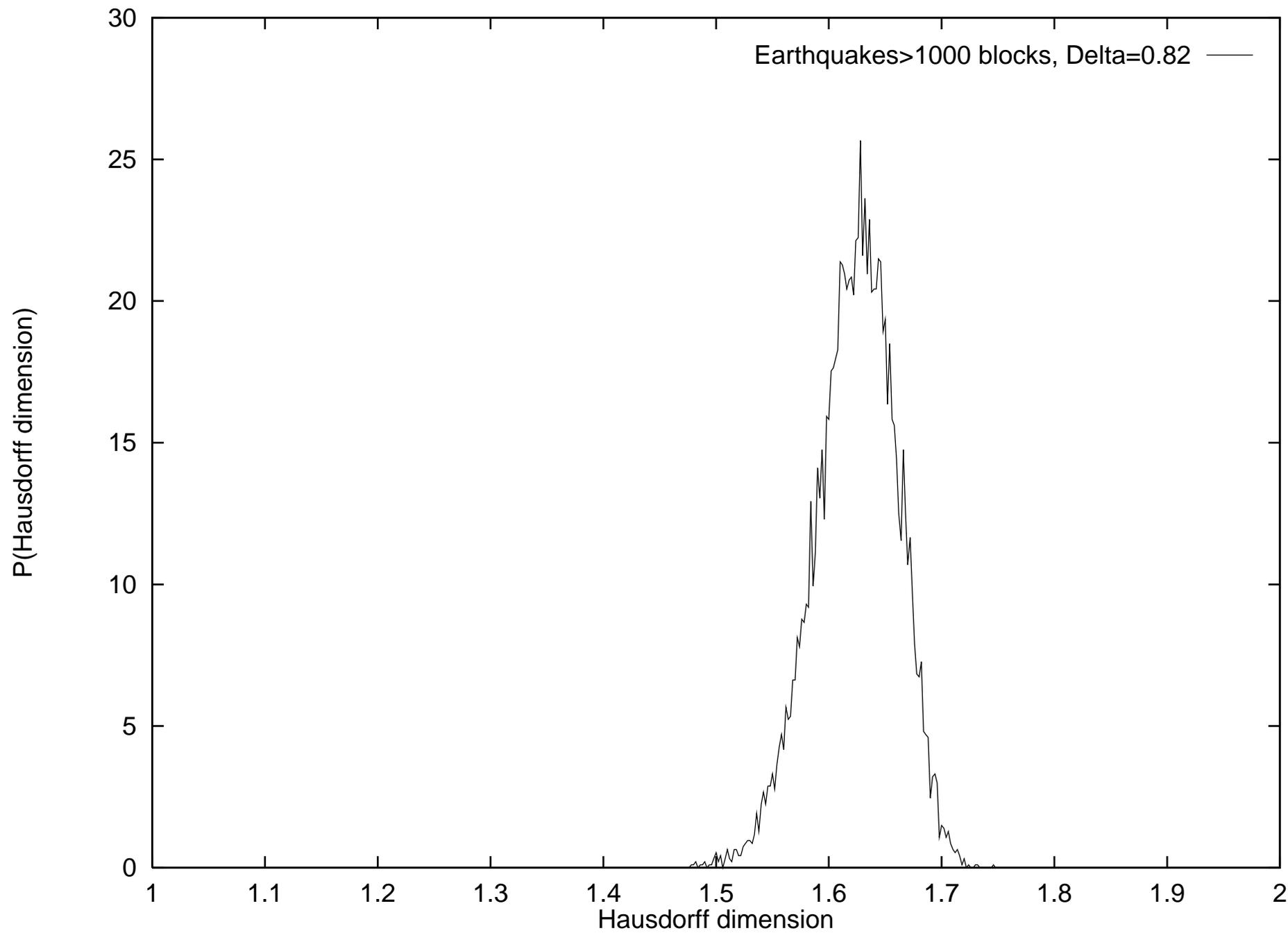
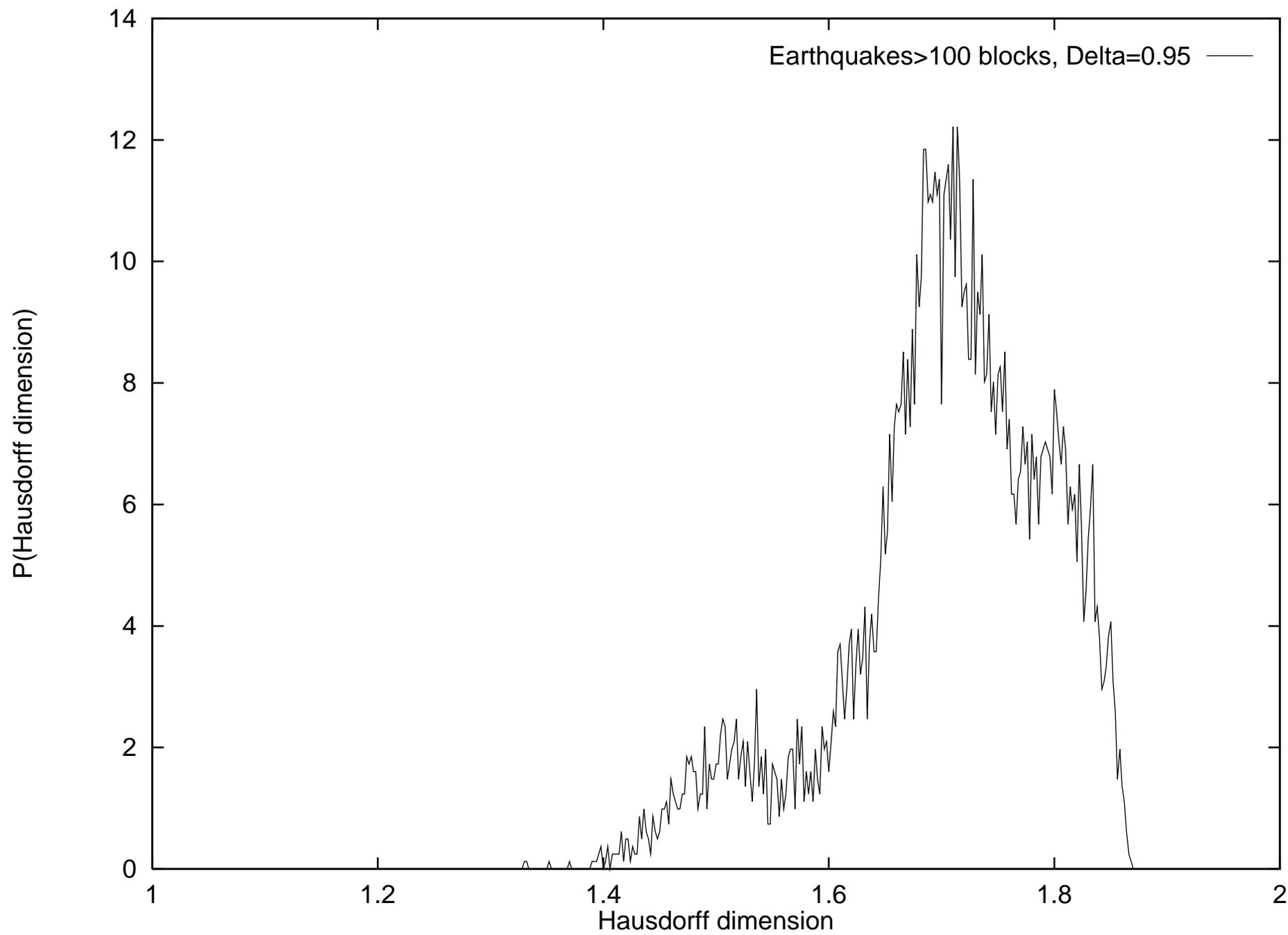
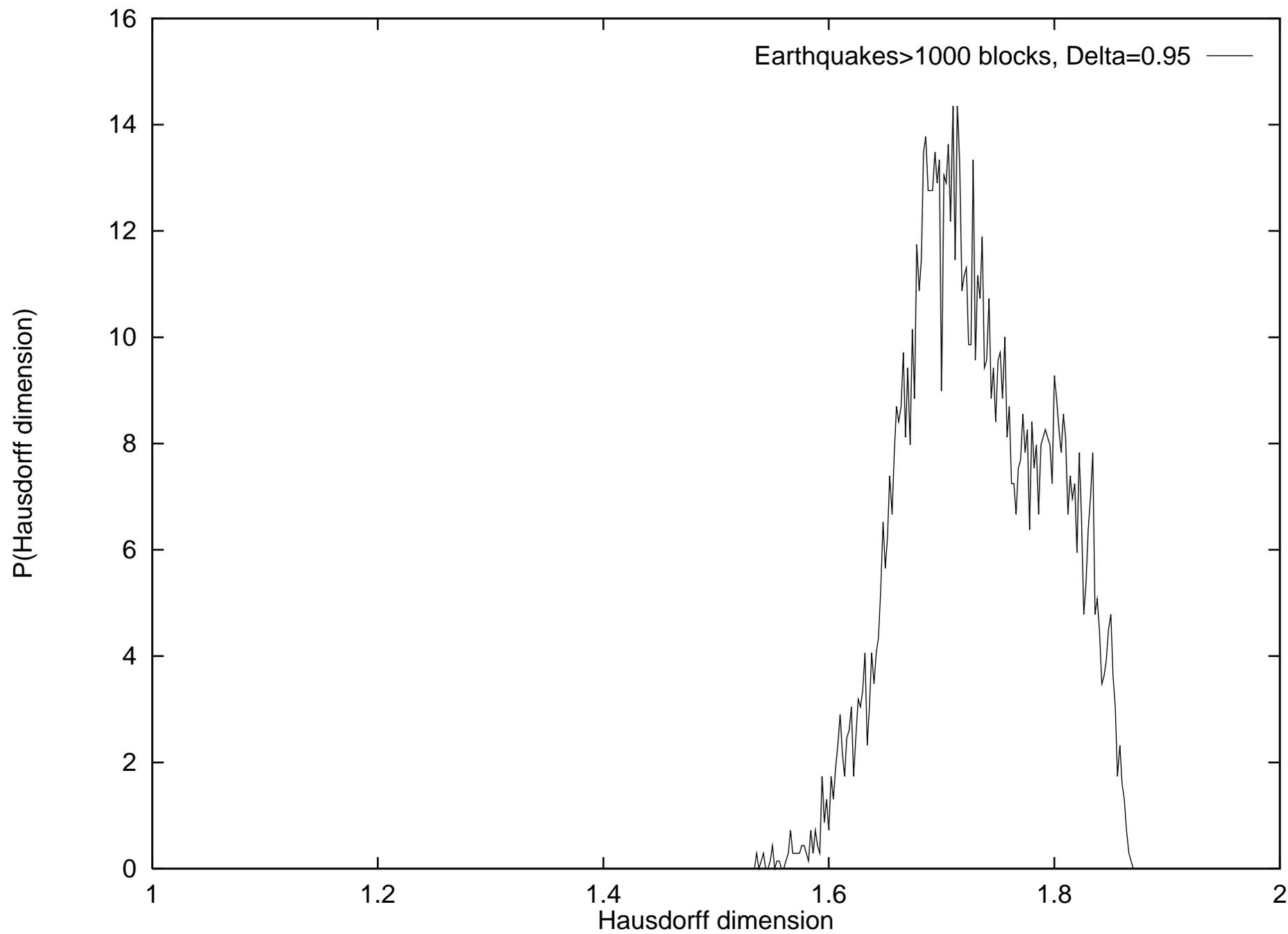
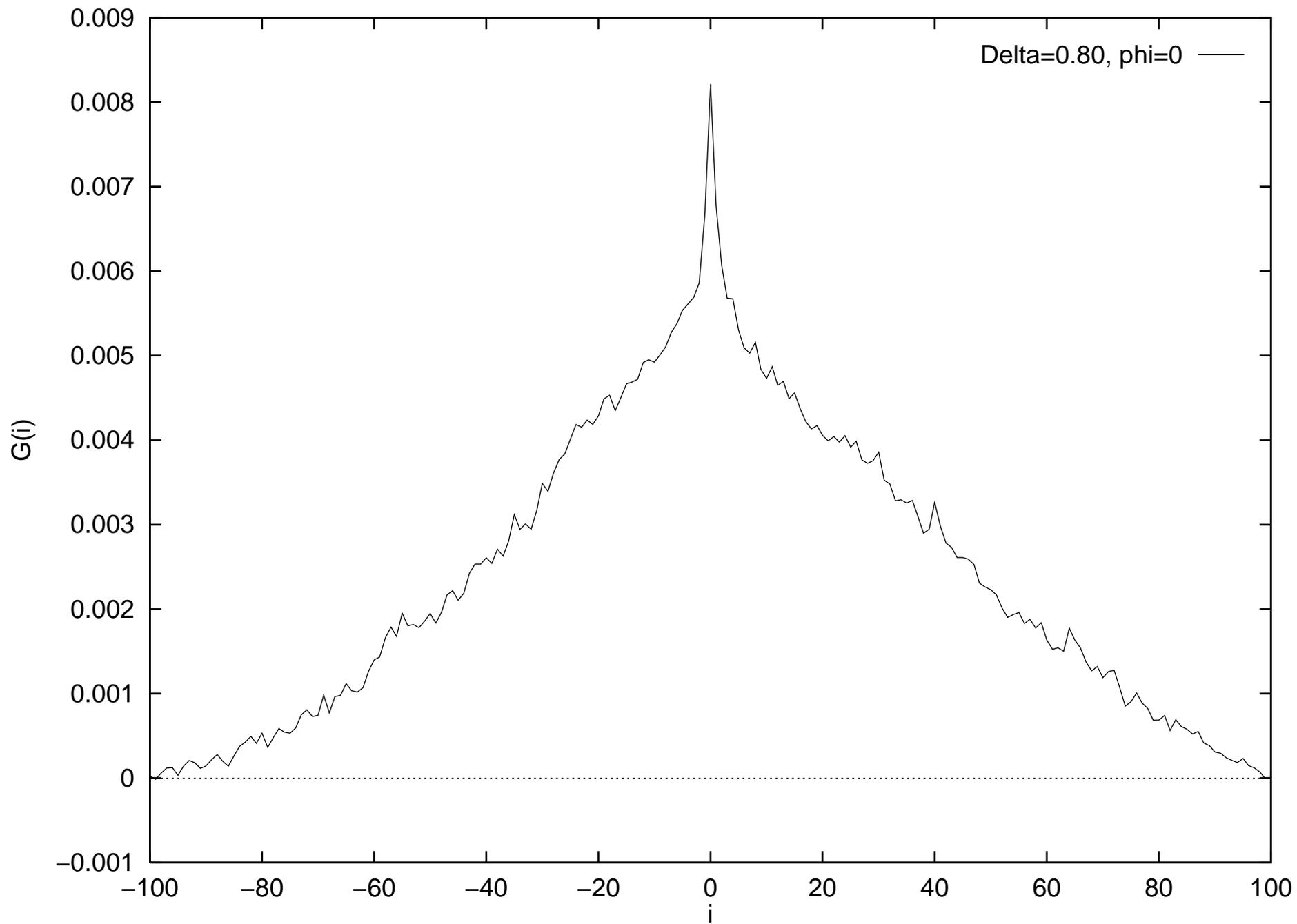
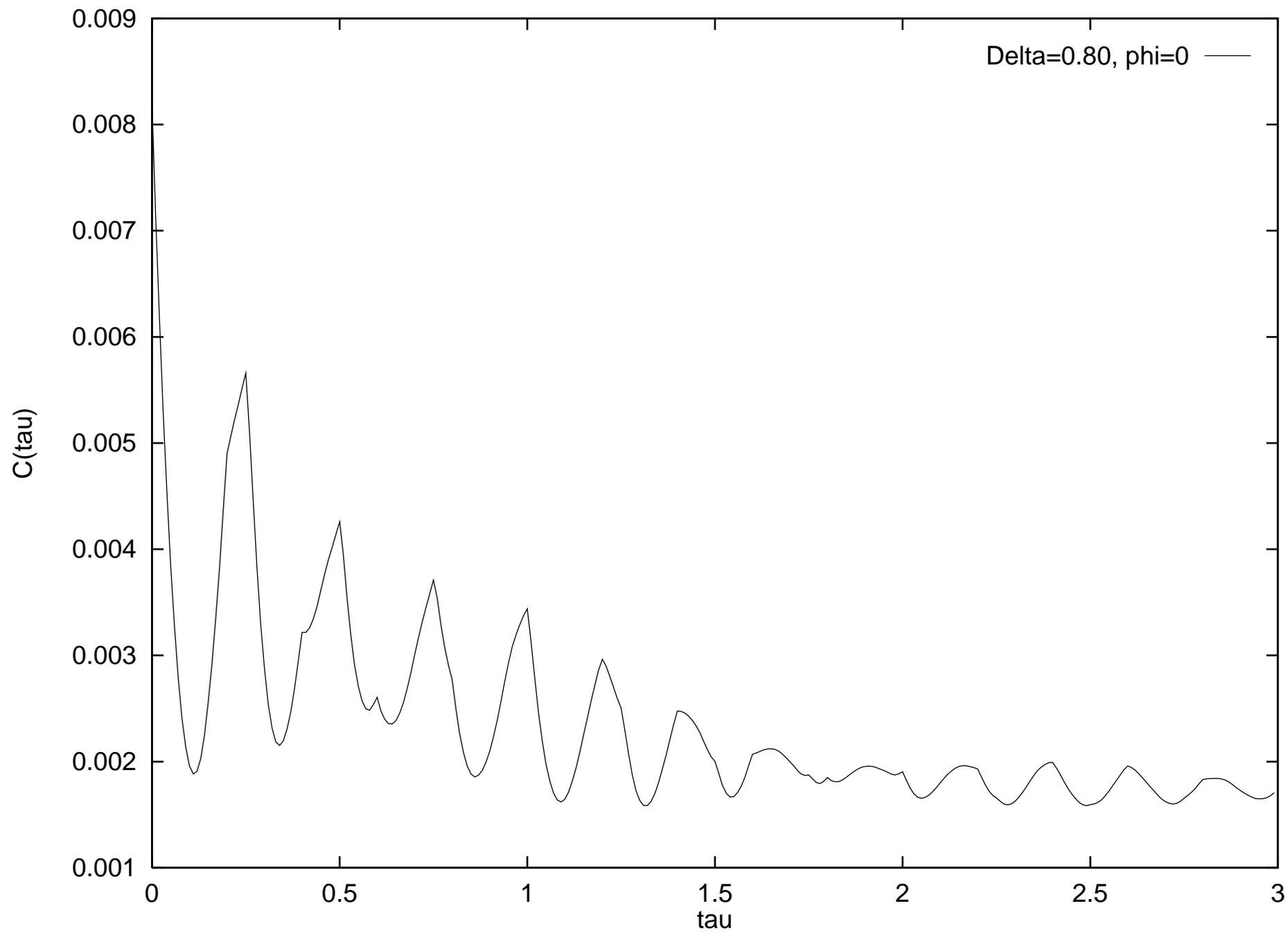
\end{document}